\documentclass[12pt,a4,prb,superscriptaddress]{article}
\usepackage{epsfig}
\title{The Moving Center of Mass of a Leaking Bob}
\author{P. Arun\footnote{e-mail:arunp92@physics.du.ac.in}\\
Department of Electronics,\\
S.G.T.B. Khalsa College\\
University of Delhi, Delhi 110 007, India.}

\begin{document}
\maketitle
\begin{abstract}
The evaluation of variation in oscillation time period of a simple
pendulum as its mass varies proves a rich source of discussion in a physics
class-room, overcoming erroneous notions carried forward by students as to 
what constitutes a pendulum's length due to picking up only the results of 
approximations and ignoring the rigorous definition. The discussion also
presents a exercise for evaluating center of mass of geometrical shapes and
system of bodies. In all, the pedagogical value of the problem is worth both
theoretical and experimental efforts. This article discusses the theoretical
considerations.  
\end{abstract}

\section{Introduction}
What happens to a simple pendulum's oscillation time period with varying
mass? This question is of pedagogical interest. An article in American
Journal of Physics \cite{ajp} addresses this issue from an experimental point 
of view, explaining variation in time period with oscillation of a burrette 
whose liquid content drips. From an introductory class point of view, the
experiment misses many important issues by using a rigid pendulum instead of
a simple pendulum. In this article, using theoretical considerations basic
ideas of defining length of the pendulum (simple as it may look, the fine
print is mostly overlooked), calculation of position of center of mass of a
body and there-after a system of two bodies etc.

The oscillation time period of a simple pendulum is given as 
\begin{eqnarray}
T_o=2 \pi \sqrt{L \over g}\label{eqn1ax}
\end{eqnarray}
where L is the length of the pendulum and g, the acceleration due to
gravity. The absence of mass term in the above expression would either imply
the incompleteness of the expression via approximations in derivation or
some oversight. A brief review of the derivation shows
\begin{eqnarray}
F=-mgsin \theta \nonumber
\end{eqnarray}
where the general expression of force is 
\begin{eqnarray}
F=m\left(dv \over dt \right)\label{eq2}
\end{eqnarray}
giving
\begin{eqnarray}
m\left(dv \over dt \right)=-mgsin \theta \nonumber
\end{eqnarray}
The linear velocity `v', can be converted to angular velocity with a useful
approximation (the small angle approximation), i.e. ${\rm sin \theta \approx
\theta}$ using the relation 
\begin{eqnarray}
v=L{d\theta \over dt}\label{eq3}
\end{eqnarray}
Thus,
\begin{eqnarray}
L\left(d^2\theta \over dt^2 \right)&=&-g \theta \nonumber\\
\left(d^2\theta \over dt^2 \right)&=&-{g \over L}=-\omega_o^2 \theta \nonumber
\end{eqnarray}
It is from ${\rm \omega_o=\sqrt{g \over L}}$, that we obtain eqn(\ref{eqn1ax}).
As per the equation, the undamped motion of the simple pendulum is indeed
mass independent. But is the above derivation rigorous and exhaustive?

\section{Time period of a Leaking Pendulum}
The above derivation innocuously drops an important definition of force, 
defined as rate of change of momentum, i.e. ${\rm F={dp \over dt}={d(mv) \over
dt}}$. Eqn(\ref{eq2}) follows only if mass is constant, which is not the
case for a leaking pendulum. Thus, the derivation would require
modifications.
\begin{eqnarray}
{d(mv) \over dt} &=& -mgsin \theta\nonumber\\
{d \over dt}\left[m{d (L\theta) \over dt}\right]  &=& -mg\theta\nonumber\\
L{d \over dt}\left[m{d \theta \over dt}\right]  &=& -mg\theta\label{eqn4a}\\
{d^2\theta \over dt}+\left({1 \over m}{dm \over dt}\right){d\theta \over dt} 
&=& -\omega_o^2\theta \label{eq4x}
\end{eqnarray}
Eqn(\ref{eq4x}) is typically that of a damped pendulum and the time period
would be given as
\begin{eqnarray}
T &=& {T_o \over \sqrt{1-{L \over 4g}\left({1 \over m}{dm \over dt}\right)^2}}
\label{eqn5a}
\end{eqnarray}
The expression typically shows how the time period would vary with variation
in mass. However, from the observations of the burette experiment as also 
from our observations in case of a pendulum made with a hollow bob filled 
with water, the time period initially increases and then starts falling.
Eqn(\ref{eqn5a}) can not explain this observation considering the rate of
change of mass of a leaking pendulum will always have values greater than or
equal to zero.  

To investigate further into the equation, we consider the length of the
pendulum also to be changing with time. Thus, the above derivation changes
from the point of eqn(\ref{eqn4a}). That is,
\begin{eqnarray}
{d(mv) \over dt} &=& -mgsin \theta\nonumber\\
{d \over dt}\left[m{d (L\theta) \over dt}\right]  &=& -mg\theta\nonumber\\
{d^2\theta \over dt^2}+
\left({2 \over L}{dL \over dt}+{1 \over m}{dm \over dt}\right)
{d\theta \over dt}
+\left({1 \over L}{d^2L \over dt^2}+{1 \over mL}{dm \over dt}{dL \over dt}
\right)
\theta &=&-\omega_o^2\theta\nonumber
\end{eqnarray}
giving an expression for time period as
\begin{eqnarray}
{2\pi \over T} &=& \sqrt{{g \over L}+ 
\left({1 \over L}{d^2L \over dt^2}+{1 \over mL}{dm \over dt}{dL \over dt}
\right)-{1 \over 4}
\left({2 \over L}{dL \over dt}+{1 \over m}{dm \over dt}\right)^2
}
\end{eqnarray}
For those who missed the rigorous definition of what constitutes the
pendulum length would ponder how the length of the string used to suspend
the bob would vary with time. That is, students carry a wrong notion that the 
length of the pendulum is the
length of the string. This, however is only true if the bob is dense and
considered a point mass with bob's radius far smaller than the length of the 
string. Practically, this is not the
case and the length of the pendulum would be length of the string and the
radius of the bob. The radius is included since the whole mass of the bob is
concentrated at its center, or its center of mass. The length of the
pendulum is hence rigorously defined as {\textsl {\textbf{distance between 
point of suspension to the center of mass of the pendulum.}}} 

In case of a leaking bob, the decreasing water would give a moving center of
mass (see fig~1). The length of the pendulum then can be written as
\begin{eqnarray}
L=l+r_o-\chi \label{eqn66}
\end{eqnarray}
where `l', `${\rm r_o}$' and `${\rm \chi}$' is the length of the string,
radius of the bob and the position of the center of mass written with
respect to the bob's center (set as origin) respectively. It is clear as
water drips, initially the center of mass moves down giving +dL/dt. At some
point when water content is low, the center of mass would tend to move back
to the center, leading to -dL/dt. This change in sign would explain
increasing time period followed by a decrease in it. The leaking pendulum
thus not only helps to illustrate the requirement to appreciate the what is
the length of the pendulum, but also adds the need for an expression of
center of mass as a function of the water level or its mass. In the passages
below we proceed to find an expression for the center of mass of our bob
consisting a shell with water.
\begin{figure}[h!!]
\begin{center}
\includegraphics[width=3in,angle=0]{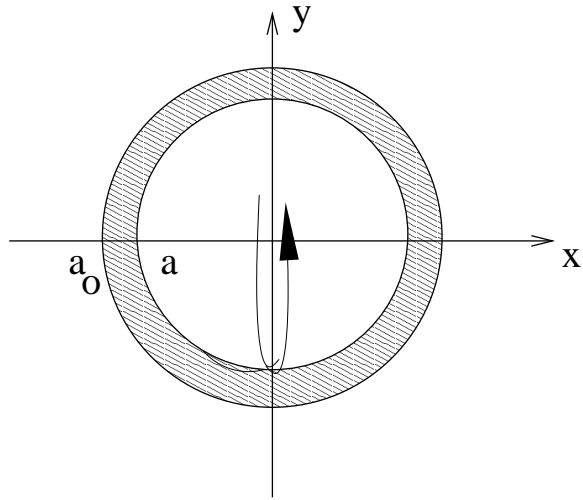}
\caption{Variation of COM with water level and resultant variation in time
period of oscillation.}  
\end{center}
\end{figure}

In the following section we derive an expression for the variable mass
simple pendulum, eseenetially considering the bob to be a hollow shell
filled with water and as it leaks the water level varies.

\section{Where is the Center of Mass}
\subsection{Of a Shell}
The calculation of the center of mass (COM) of a body starts with a
evaluation of the body's mass. Since the shell in question has spherical 
symmetry, the integral to calculate the mass is best done in polar
coordinates. Thus,
\begin{eqnarray}
m_{shell} &=& \rho_{shell}\int_{r=a}^{a_o} \int_{\theta=0}^{2 \pi}
\int_{\phi=-\pi/2}^{\pi/2}
dr \times r d\theta \times rcos \phi d\phi  \nonumber\\
&=& \rho_{shell}\int_{r=a}^{a_o} \int_{\theta=0}^{2 \pi} 
\int_{\phi=-\pi/2}^{\pi/2}
r^2 cos\phi drd\theta d\phi  \nonumber\\
&=& {4 \over 3} \pi \rho_{shell}(a_o^3-a^3)\label{mshell}
\end{eqnarray}
where `${\rm a_o}$' and `a' are the outer and inner radius of the shell
whose density is `${\rm \rho_{shell}}$'. The 
general expression for COM is given as
\begin{eqnarray}
\chi={\Sigma m_ir_i \over M}\label{eq1}
\end{eqnarray}
where M is the total mass of the body and ${\rm m_i}$ is the mass of small
volume (${\rm dx_idy_idz_i}$) of the body at ${\rm r_i}$ from the origin.
Applying this to the problem of shell (using eqn \ref{mshell}), we have 

\begin{eqnarray}
&=&{\rho_{shell} \over m_{shell}}\int_a^{a_o}
\int_0^{2\pi}\int_{-\pi/2}^{\pi/2} r^2cos\phi 
drd\theta d\phi (rcos\theta cos \phi \hat i+
rsin\theta cos \phi \hat j+ r sin \phi \hat k)\nonumber\\
&=&{\rho_{shell}(a_o^4-a^4) \over 4m_{shell}}
\int_0^{2\pi}\int_{-\pi/2}^{\pi/2} cos\phi 
d\theta d\phi (cos\theta cos \phi \hat i+
sin\theta cos \phi \hat j+ sin \phi \hat k)\nonumber\\
&=&{\rho_{shell}(a_o^4-a^4) \over 4m_{shell}}
\int_0^{2\pi}\int_{-\pi/2}^{\pi/2} (cos\theta cos^2 \phi \hat i+
sin\theta cos^2 \phi \hat j+ sin \phi cos \phi \hat k)d\theta d\phi\nonumber\\
&=&{\rho_{shell}(a_o^4-a^4) \over 4m_{shell}}
\left[0\hat i+ 0\hat j+ 2\pi \int_{-\pi/2}^{\pi/2} 
(sin \phi cos \phi d\phi) \hat k\right]\nonumber\\
&=& 0\hat i + 0\hat j+0\hat k\label{com2} 
\end{eqnarray}
The COM is at the center of the shell.

\subsection{Water Body}
The water mass filled in the shell, when the shell is filled can be
considered to be a sphere of radius `a', the shell's inner radius. Now, 
consider disc of thickness `dz' is cut from the sphere at a
distance `z' from the center. The disc has radius `r' (see fig 2) and hence 
it's area would be 
\begin{eqnarray}
A=\pi r^2\nonumber
\end{eqnarray}
The disc volume would be
\begin{eqnarray}
dV=\pi r^2 dz\nonumber
\end{eqnarray}
\begin{figure}[h!!]
\begin{center}
\includegraphics[width=3in,angle=0]{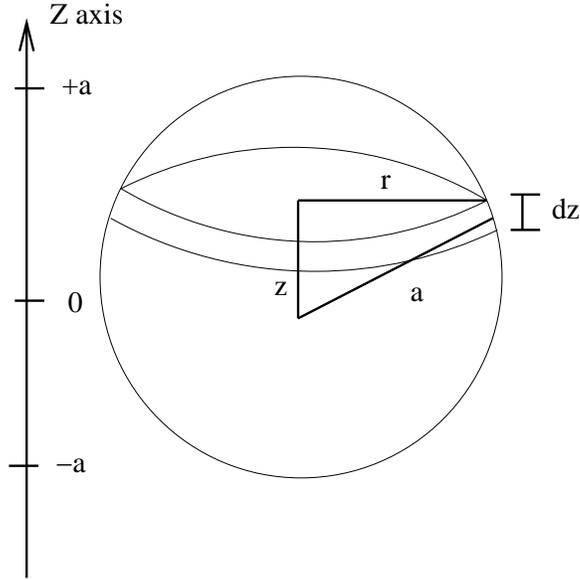}
\caption{For deriving the center of mass of water body, we can consider the
body to be sum of pile of disc of thickness `dz' at a distance `z' from the
center. The disc radius `r' would be a function of the distance of the disc
from the center.}  
\end{center}
\end{figure}

The mass associated with this disc would be 
\begin{eqnarray}
dM &=& \rho_{water} dV\nonumber\\
&=& \rho_{water} \pi r^2 dz
\end{eqnarray}
The net mass of the `water sphere' hence would be
\begin{eqnarray}
M_{water} &=& \rho_{water} \pi\int_{-a}^a r^2dz\nonumber
\end{eqnarray}
But radius of disc would depend on how far away from the center is the disc
cut, hence ${\rm r \rightarrow r(z)}$, which is obtained from simple rule
${\rm r^2=a^2-z^2}$. Hence,
\begin{eqnarray}
M_{water} &=& \rho_{water} \pi\int_{-a}^a (a^2-z^2)dz\nonumber\\
&=& \rho_{water} \pi \left(a^2z-{z^3 \over 3}\right)\Biggr\vert_{-a}^a\nonumber\\
&=& \rho_{water} \pi \left[\left(a^3-{a^3 \over 3}\right)-
\left(-a^3+{a^3 \over 3}\right)\right]\nonumber\\
&=& \rho_{water} \left({4\pi a^3 \over 3}\right)\nonumber
\end{eqnarray}
In case the sphere is not completely filled then the calculations remain the
same, however, the limits change. Say the water level is at `h', the limits
change and calculations proceed as
\begin{eqnarray}
M_{water} &=& \rho_{water} \pi\int_{-a}^h (a^2-z^2)dz\nonumber\\
&=& \rho_{water} \pi \left(a^2z-{z^3 \over 3}\right)\Biggr\vert_{-a}^h\nonumber\\
&=& \rho_{water} \pi \left[\left(a^2h-{h^3 \over 3}\right)-
\left(-a^3+{a^3 \over 3}\right)\right]\nonumber\\
&=& \rho_{water} \pi \left({2a^3 \over 3}+a^2h-{h^3 \over 3}\right)\label{eq24}
\end{eqnarray}

The position of the center of mass then is calculated by solving the 
following integral
\begin{eqnarray}
\chi_{water} &=& {\rho_{water} \pi \over M_{water}}
\int_{-a}^h z(a^2-z^2)dz\nonumber\\
&=& {\rho_{water} \pi \over M_{water}}\left({a^2z^2 \over 2}-
{z^4 \over 4}\right)
\Biggr\vert_{-a}^h\nonumber\\
&=& {\rho_{water} \pi \over M_{water}} 
\left[\left({a^2h^2 \over 2}-{h^4 \over 4}\right)-
\left({a^4 \over 2}-{a^4 \over 4}\right)\right]\nonumber\\
&=& {\rho_{water} \pi \over M_{water}} \left(-{a^4 \over 4}+{a^2h^2 \over 2}-
{h^4 \over 4}\right)\label{eq25}
\end{eqnarray}

\subsection{Of Leaking Bob}
The COM of the pendulum's bob made by a thin shell filled with water can be
evaluated using standard formula
\begin{eqnarray}
\chi_{bob}={m_{shell}\chi_{shell}+M_{water}\chi_{water} 
\over m_{shell}+M_{water}}\label{com1}
\end{eqnarray}
As evaluated in eqn(\ref{com2}), the shell's COM will always be at it's
center, which we take as the origin. Hence, eqn(\ref{com1}) reduces to
\begin{eqnarray}
\chi_{bob}={M_{water}\chi_{water}
 \over m_{shell}+M_{water}}\label{com4}
\end{eqnarray}

Using eqn(\ref{mshell}), eqn(\ref{eq24}) and eqn(\ref{eq25}) we have
\begin{eqnarray}
\chi_{bob}&=&{\rho_{water} \pi \left(-{a^4 \over 4}+{a^2h^2 \over 2}-
{h^4 \over 4}\right)
\over {4 \over 3} \pi \rho_{shell}(a_o^3-a^3)+
\rho_{water} \pi \left({2a^3 \over 3}+a^2h-{h^3 \over 3}\right)}\nonumber\\
&=&{\rho_{water} \left(-{a^4 \over 4}+{a^2h^2 \over 2}-
{h^4 \over 4}\right)
\over {4 \over 3} \rho_{shell}(a_o^3-a^3)+
\rho_{water} \left({2a^3 \over 3}+a^2h-{h^3 \over 3}\right)}\nonumber\\
&=&{3\rho_{water} (-a^4+2a^2h^2-h^4)
\over 16 \rho_{shell}(a_o^3-a^3)+
4\rho_{water} (2a^3+3a^2h-h^3)}\nonumber
\end{eqnarray}
The above equation will give the variation of bob's COM as water leaks. The
density of water is unity, hence 
\begin{eqnarray}
\chi_{bob} ={3 (-a^4+2a^2h^2-h^4) \over 16 \rho_{shell}(a_o^3-a^3)+
4 (2a^3+3a^2h-h^3)}\label{com5}
\end{eqnarray}
For the purpose of plotting it would be better to define a normalized
variable, (h/a) and rewrite the above equation. We have
\begin{eqnarray}
\chi_{bob} &=&{3a^4 \left[-1+2\left({h \over a}\right)^2-\left({h\over
a}\right)^4\right] \over 16
\rho_{shell}a^3\left[\left({a_o\over a}\right)^3-1\right]+
4a^3 \left[2+3\left({h\over a}\right)-
\left({h\over a}\right)^3\right]}\nonumber\\
&=&{3a (-1+2x^2-x^4) \over 16
\rho_{shell}\left[\left({a_o\over a}\right)^3-1\right]+
4(2+3x-x^3)}\label{com6}
\end{eqnarray}
where ${\rm {a_o \over a}}$ would have a value greater than unity. The first
term in the denominator would depend on the shell's thickness and it's
density, hence we write
\begin{eqnarray}
\chi_{bob} &=&{3a (-1+2x^2-x^4) \over K+4(2+3x-x^3)}\label{com7}
\end{eqnarray}
\begin{figure}[h!!]
\begin{center}
\includegraphics[width=2.1in,angle=-90]{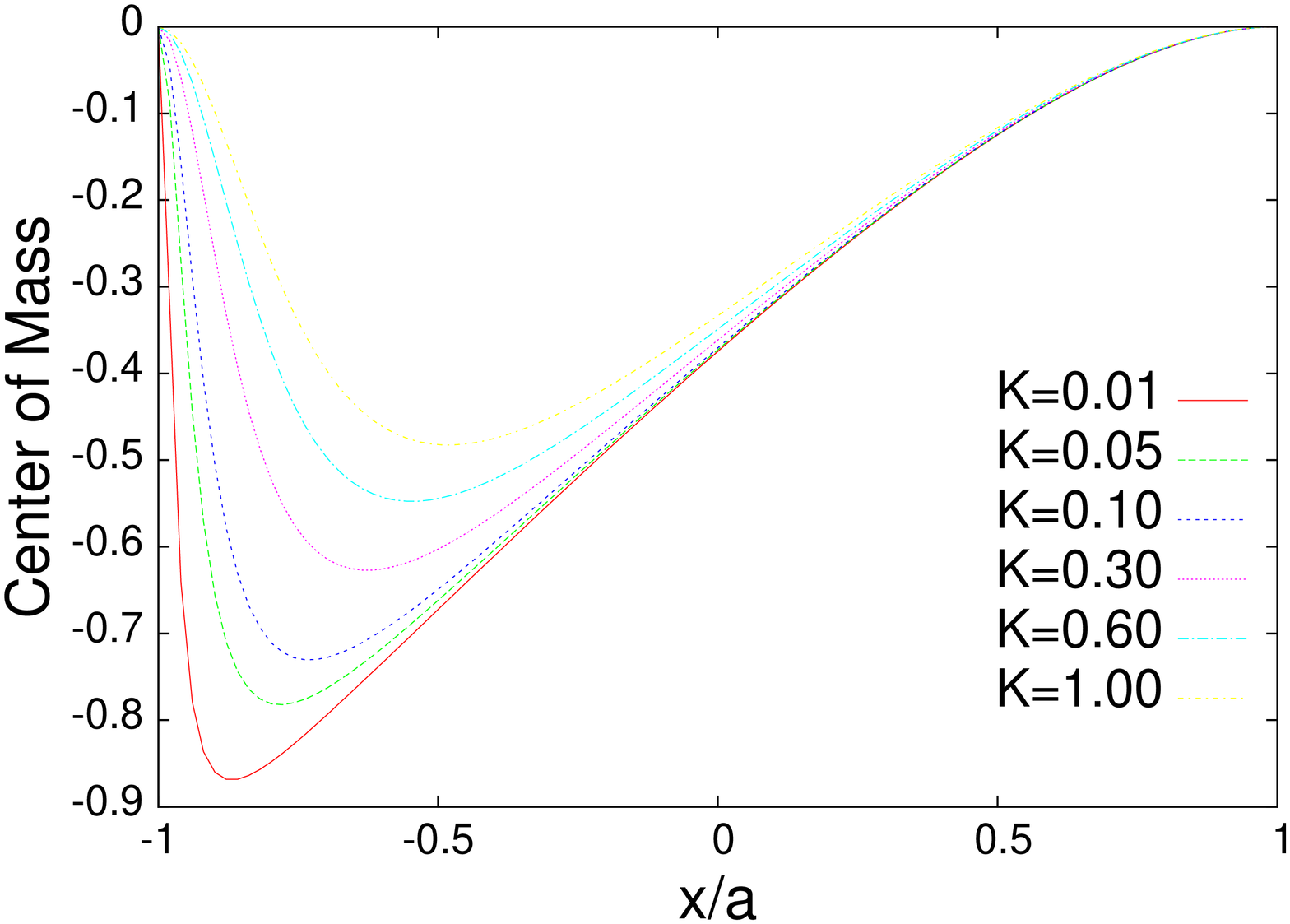}
\hfil
\includegraphics[width=2.1in,angle=-90]{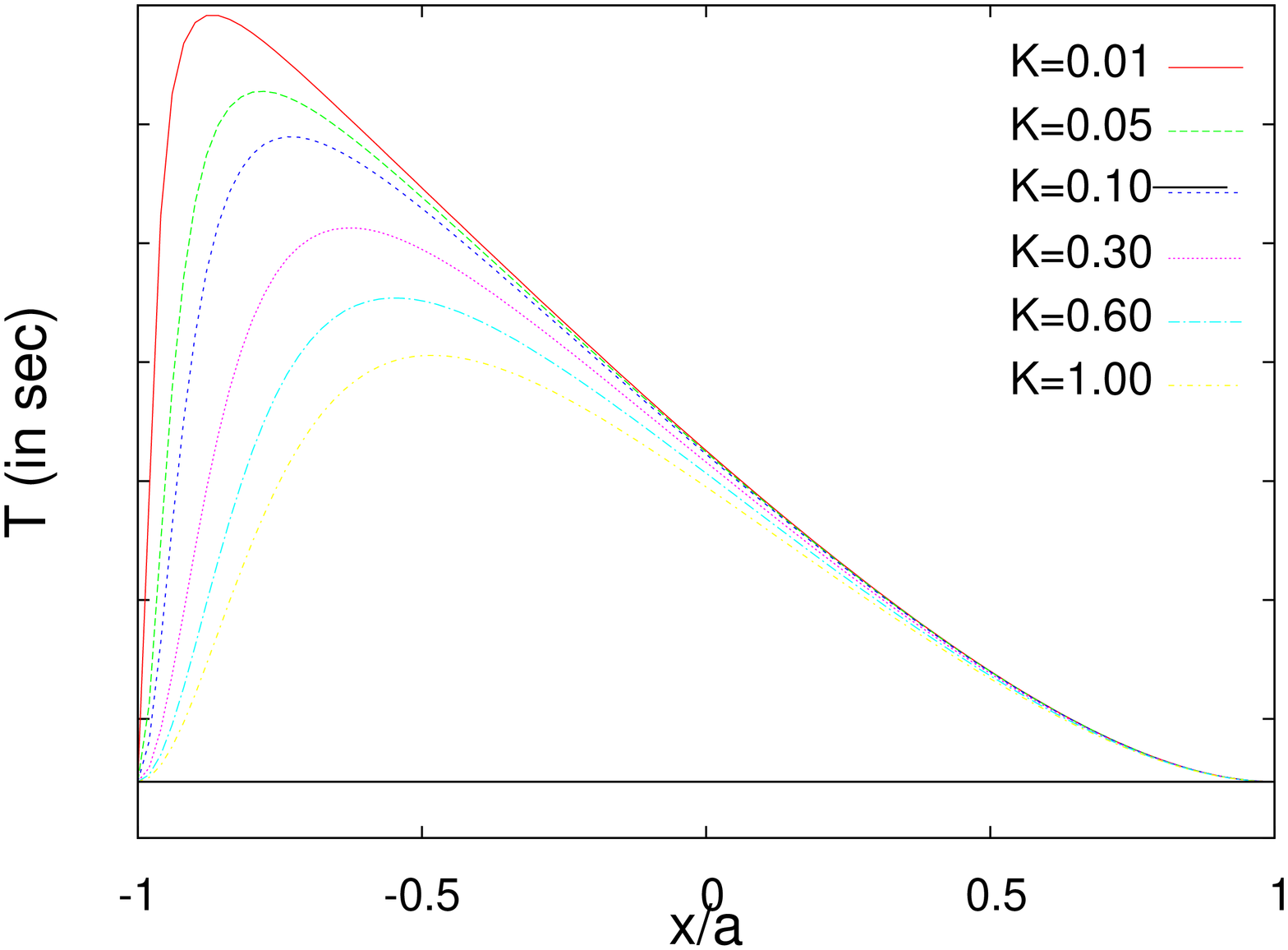}
\caption{Variation of COM with water level and resultant variation in time
period of oscillation.}  
\end{center}
\end{figure}

The above mathematics imply that the varying mass of the pendulum (as water
flows from the bob) results in change in center of mass and in turn the 
length of the pendulum (basically from results flowing from eqn \ref{eqn66},
eqn \ref{eq24} and eqn \ref{com7}). In other words the length of a varying mass
pendulum is closely related to the mass. Figure~3 shows the changing position 
of the bob's center of mass as the water level in it changes along with
resulting oscillation time period. The curves were 
generated using eqn(\ref{com7}), eqn(\ref{eqn66}) and eqn(\ref{eqn1ax}).
Though not an exhaustive calculation it easily shows the increase followed
by decrease in time period of oscillation as the pendulum's mass varies.

The plots are a family of curves, generated for various
``shell parameters (K)'', which depends of the shell material's density and
it's thickness. It is clear from these curves that to get good resolvable
experimental results, it is best to use a shell of very small thickness and
moderate density (`K' small). This ensures the center of mass of the
shell-water body system is strongly controlled by rhe water body. 
Based on these ideas experimental results are 
being gathered and would be reported in future. 



\section{Conclusion}
The article discusses the problem of a variable mass pendulum. The
increasing followed by decreasing time period of oscillation is a novel
feature. Though simply explained via variation in the pendulum length, the
discussion of this problem in a classroom is of rich pedagogical value.
Experimental verification of these ideas would further enrich the experience
of applying concepts such as center of mass which is usually studied with 
theoretical emphasis.

\end{document}